\documentclass[12pt]{article}
\usepackage[utf8]{inputenc}
\usepackage{amsmath}
\usepackage{amssymb}
\usepackage{mathrsfs}
\usepackage{graphicx}
\usepackage{cite}
\usepackage{caption}
\usepackage{subcaption}
\usepackage{epstopdf}
\usepackage{float}
\usepackage[makeroom]{cancel}
\usepackage[margin=1.2in]{geometry}
\usepackage{color}

\begin{document}

\begin{titlepage}

%\begin{center}
%{\hbox to\hsize{
%\hfill \bf hep-ph/??? }}
{\hbox to\hsize{\hfill Feb. 2017 }}

\bigskip \vspace{3\baselineskip}

\begin{center}
{\bf \large
Probing the MSSM explanation of the muon g-2 anomaly in dark matter experiments and at a 100 TeV $pp$ collider}

\bigskip

\bigskip

{\bf Archil Kobakhidze$^1$, Matthew Talia$^1$ and Lei Wu$^{1,2}$ \\ }

\smallskip

{ \small \it
$^1$ ARC Centre of Excellence for Particle Physics at the Terascale, \\
School of Physics, The University of Sydney, NSW 2006, Australia \\
$^2$ Department of Physics and Institute of Theoretical Physics, Nanjing Normal University, \\ Nanjing, Jiangsu 210023, China\\
E-mails: archil.kobakhidze, matthew.talia, lei.wu1@sydney.edu.au
\\}

\bigskip

\bigskip

\bigskip

{\large \bf Abstract \\ }
\end{center}
\noindent
We explore the ability of current and future dark matter and collider experiments in probing anomalous magnetic moment of the muon, $(g-2)_\mu$, within the Minimal Supersymmetric Standard Model (MSSM). We find that the latest PandaX-II/LUX-2016 data gives a strong constraint on parameter space that accommodates the $(g-2)_{\mu}$ within $2\sigma$ range, which will be further excluded by the upcoming XENON-1T (2017) experiment. We also find that a 100 TeV $pp$ collider can cover most of our surviving samples that satisfy DM relic density within $3\sigma$ range through $Z$ or $h$ resonant effect by searching for trilepton events from $\tilde{\chi}^0_2\tilde{\chi}^+_1$ associated production. While the samples that are beyond future sensitivity of trilepton search at a 100 TeV $pp$ collider and the DM direct detections are either higgsino/wino-like LSPs or bino-like LSPs co-annihilating with sleptons. Such compressed regions may be covered by the monojet(-like) searches at a 100 TeV $pp$ collider.
\end{titlepage}

\section{Introduction}
The discovery of Higgs boson \cite{higgs-atlas,higgs-cms} and subsequent measurements of its properties completed the Standard Model (SM) and provided it with very convincing evidence for the simplest perturbative realisation of the electroweak symmetry breaking (EWSB). Despite this overwhelming empirical success, our understanding of EWSB is incomplete. Namely, the quantum corrections are known to drive the Higgs mass (and hence the electroweak scale) towards high-energy scales and thus the SM requires unnaturally precise fine-tuning of parameters to satisfy the observations. In addition, observations of neutrino oscillations and dark matter (DM) certainly require beyond the standard model (BSM) physics.

Another deviation from the SM prediction are long seen in the measurements of the anomalous magnetic moment of the muon, $a_{\mu}=(g-2)_{\mu}/2$ \cite{g2-review-1,g2-review-2,g2-review-3,g2-review-4}. The recently measured value \cite{g2-exp},
\begin{align}
\Delta a_\mu^{\text{Exp$-$SM}}&=
\begin{cases}
(28.7 \pm 8.0 ) \times 10^{-10} \mbox{\cite{g2-sm-1}}, \\
(26.1 \pm 8.0 ) \times 10^{-10} \mbox{\cite{g2-sm-2}}~,
\end{cases}
\label{deviation}
\end{align}
are more than $3\sigma$ away from the SM prediction,
which includes improved QED \cite{g2-sm-qed} and electroweak \cite{g2-sm-ew} contributions. The upcoming experiments at NBL will measure the $(g-2)_\mu$ with a precision of 0.14 ppm \cite{g2-exp-new}, which would potentially allow a $5\sigma$ discovery of new physics through such measurements. Needless to say, there are several candidate explanations for $(g-2)_\mu$ anomaly proposed within various new physics frameworks.

The weak-scale supersymmetry (SUSY) has long been the dominant paradigm for new particle physics. The minimal supersymmetric standard model (MSSM) not only provides an elegant solution to the hierarchy problem but also may successfully explain $(g-2)_{\mu}$ anomaly \cite{g2-mssm-0,g2-mssm-2,g2-mssm-3,g2-mssm-6,g2-mssm-10,g2-mssm-15,g2-mssm-16,g2-mssm-17,g2-mssm-19,g2-mssm-20,g2-mssm-21,g2-mssm-22,g2-mssm-23,g2-mssm-24,g2-mssm-25,g2-mssm-26,g2-mssm-27,g2-mssm-28,g2-mssm-29,g2-mssm-30,g2-mssm-31}. In the MSSM, the most significant contribution to $a_{\mu}$ is due to the one-loop diagrams involving the smuon $\widetilde{\mu}$, muon sneutrino $\widetilde{\nu}_{\mu}$, neutralinos $\widetilde{\chi}^{0}$ and charginos $\widetilde{\chi}^{\pm}$. The one-loop contribution to $a_\mu$ arises if there is a chirality flip between incoming and outgoing external muon lines, which may be induced through the $L-R$ mixing in the smuon sector or the SUSY Yukawa couplings of Higgsinos to muon and $\widetilde{\mu}$ or $\widetilde{\nu}_{\mu}$. Therefore, these contributions to $a_\mu$ are typically proportional to $m^2_\mu/M^2_{SUSY}$. Thus, to generate the sizable contributions to $a_\mu$, the SUSY scale $M_{SUSY}$ encapsulating slepton and electroweakino masses has to be around $O(100)$ GeV. So, the detection of light sleptons and electroweakinos will provide a test for the MSSM solution to the $(g-2)_{\mu}$ problem.

The negative results of direct searches for sparticles during the LHC Run-1 have pushed up the mass limits of the first two generation squarks and gluino into the TeV region \cite{lhc-gluino-atlas,lhc-gluino-cms}. The third generation squarks have been tightly constrained in the simplified models \cite{lhc-stop-atlas,lhc-stop-cms}, such as in Stealth SUSY \cite{lhc-steath} and Natural SUSY \cite{lhc-ns-1,lhc-ns-2,lhc-ns-3,lhc-ns-4,lhc-ns-5,lhc-ns-6}. Unlike the colored sparticles, the bounds on the sleptons \cite{lhc-slepton-atlas,lhc-slepton-cms} and electroweakinos \cite{lhc-ewkino-atlas,lhc-ewkino-cms} are relatively weak, especially for the region of compressed spectrum.
The lightest neutralino still remains as a successful dark matter (DM) candidate and significant effort has been made to obtain a lower mass limit on the neutralino LSP in MSSM, see e.g. \cite{mssm-dm-1,mssm-dm-2,mssm-dm-3,mssm-dm-4}.

In this paper, we explore the potential of the current and future dark matter and collider experiments to probe the anomalous magnetic moment of the muon within the MSSM. Using LEP and Higgs data and demanding that the theory accommodates
$(g-2)_\mu$ measurements within $2\sigma$ range, we derive bounds on the electroweakino masses. Following this, we impose dark matter constraints from Planck, PandaX-II/LUX 2016 data and constraints from LHC searches for dilepton and trilepton events. Then, we evaluate the prospect of a future 100 TeV hadron collider in probing electroweakinos in trilepton events within this scenario. Finally, our conclusions are presented.

\section{$(g-2)_\mu$ in MSSM} \label{g-2}
The low-energy effective operator for magnetic dipole moment (MDM) is given by:
\begin{equation}
{\cal L}_{MDM}=\frac{e}{4m_\mu}a_{\mu}\bar{\mu}\sigma_{\rho\lambda}\mu F^{\rho\lambda}.
\end{equation}
where $e$ is the electric charge and $m_\mu$ is the muon mass. $F^{\rho\lambda}$ is the field strength of the photon field and $\sigma_{\rho\lambda}=\frac{i}{2}[\gamma_\rho, \gamma_\lambda]$.

\begin{figure}[t]
	\centering
    \includegraphics[width=0.45\textwidth]{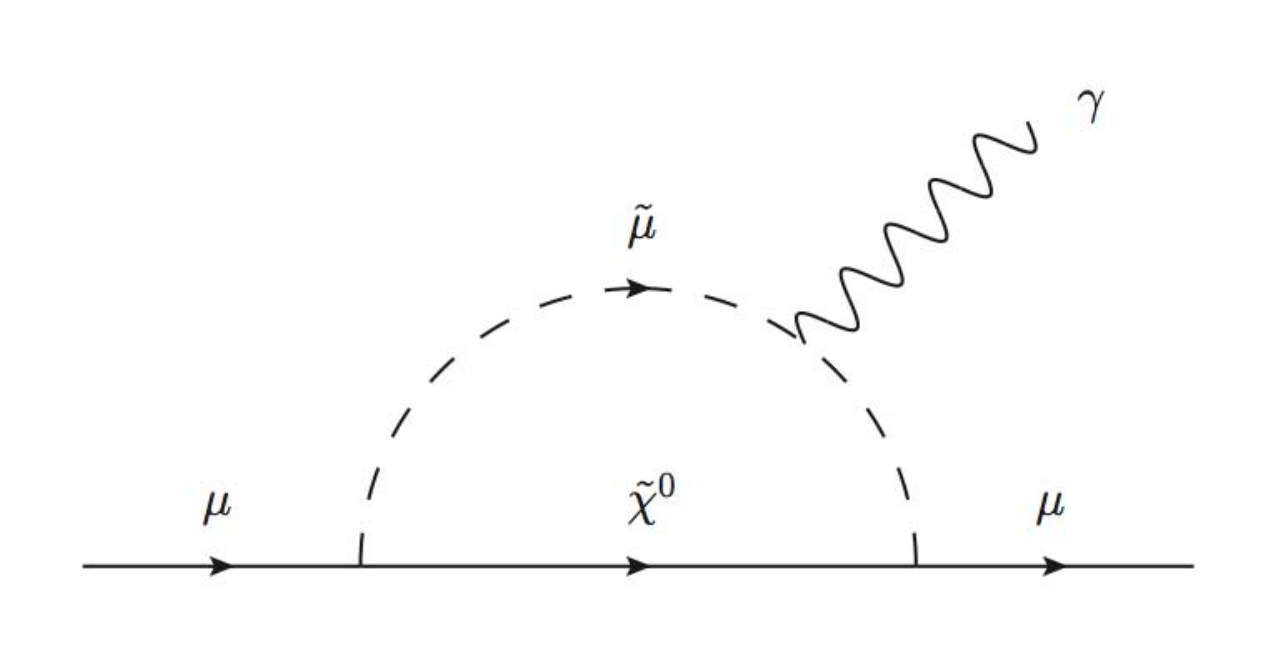}
    \includegraphics[width=0.45\textwidth]{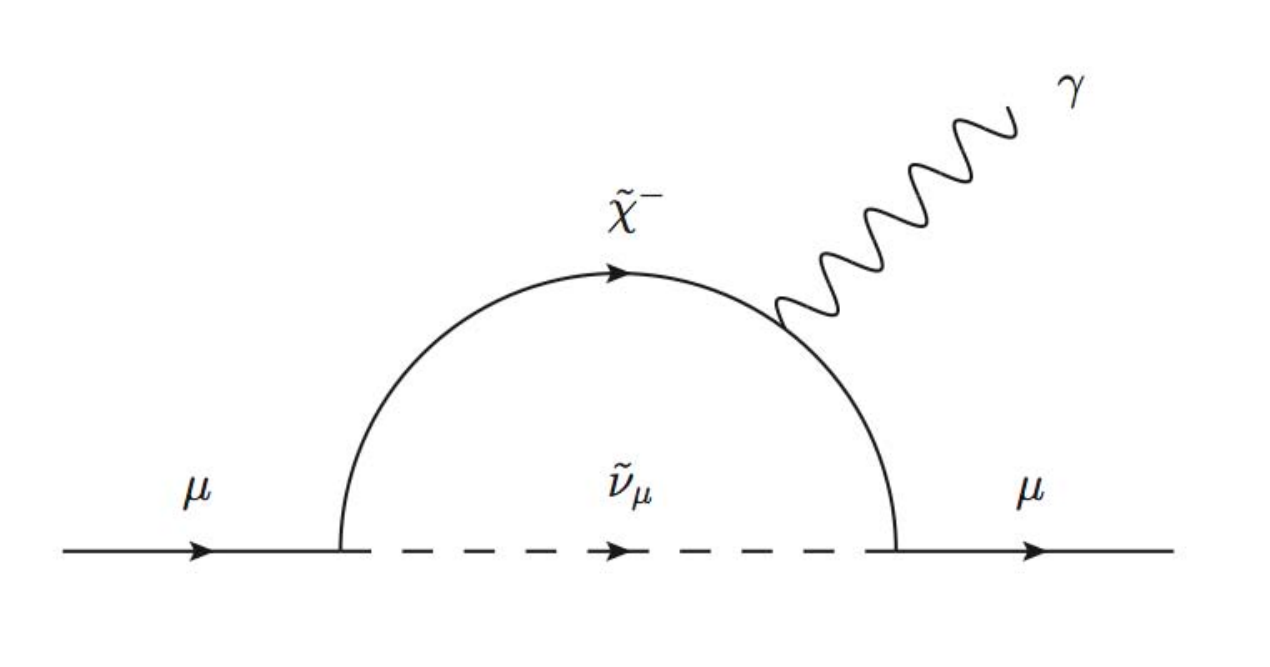}
	\caption{One-loop diagram contributions of the MSSM to the muon anomalous magnetic moment, $(g-2)_{\mu}$. The first involves a smuon-neutralino (left) and the second a chargino-muon sneutrino loop (right).}
	\label{g2diagram}
\end{figure}
In the MSSM, there are essentially two types of diagrams which contribute to $a_\mu$ at one-loop, i.e. one is the $\widetilde{\chi}^0 - \widetilde{\mu}$ loop diagram (left panel of Fig.~\ref{g2diagram}) and the other is the chargino $\widetilde{\chi}^\pm - \widetilde{\nu}_\mu$ loop diagram (right panel of Fig.~\ref{g2diagram}). The expressions for one-loop SUSY corrections to $a_{\mu}$ (including the complex phases effects) are given by \cite{g2-mssm-2}
\begin{eqnarray}
a_{\mu}^{\widetilde\chi^0}&=&\frac{m_{\mu}}{16\pi^2}\sum_{i,\alpha}\Big\{-\frac{m_{\mu}}{12 m^2_{\widetilde\mu_m}}(|n^L_{i\alpha}|^2+|n^R_{i\alpha}|^2)F_1^N(x_{i\alpha})
+\frac{m_{\widetilde\chi_i^0}}{3m^2_{\widetilde\mu_m}}\text{Re}(n^L_{i\alpha}n^R_{i\alpha})F_2^N(x_{i\alpha})\Big\},\label{eq:a2-1}\\
a_{\mu}^{\widetilde\chi^+}&=&\frac{m_{\mu}}{16\pi^2}\sum_{j}\Big\{\frac{m_{\mu}}{12 m^2_{\widetilde\nu_{\mu}}}(|c_j^L|^2+|c_j^R|^2)F_1^C(x_j)
+\frac{2m_{\widetilde\chi_j^{\pm}}}{3m^2_{\widetilde\nu_{\mu}}}\text{Re}(c_j^Lc_j^R)F_2^C(x_j)\Big\}
\label{eq:a2-2}
\end{eqnarray}
where $i=1,2,3,4$, $j=1,2$ and $\alpha=1,2$ denotes the neutralino, chargino and smuon mass eigenstates, respectively. The couplings are defined as
\begin{eqnarray}
n^R_{i\alpha}&=& \sqrt{2}g_1 N_{i1}X_{\alpha2}+y_{\mu}N_{i3}X_{\alpha1}, \nonumber \\
n^L_{i\alpha}&=& \frac{1}{\sqrt{2}}(g_2 N_{i2}+g_1 N_{i1})X^*_{\alpha1}-y_{\mu}N_{i3}X^*_{\alpha2}, \nonumber \\
c^R_j&=& y_{\mu}U_{j2}, \nonumber \\
c^L_j&=&-g_2 V_{j1},
\label{eq:c}
\end{eqnarray}
where the muon Yukawa coupling $y_{\mu}=g_2 m_{\mu}/\sqrt{2} m_W \cos\beta$. $N$ are the neutralino and $U,V$ are the chargino mixing matrices, respectively. $X$ denotes the slepton mixing matrix. In terms of the kinematic variables $x_{i\alpha}=m^2_{\widetilde\chi_i^0}/m^2_{\widetilde\mu_\alpha}$ and $x_j=m^2_{\widetilde\chi_j^{\pm}}/m^2_{\widetilde\nu_{\mu}}$, the loop functions $F$ are defined as follows
%%%%%%%%%%%%%%%%%%%
\begin{eqnarray}
F_1^N(x)&=&\frac{2}{(1-x)^4}\Big[1-6x+3x^2+2x^3-6x^2 \ln x\Big], \nonumber \\
F_2^N(x) &=&\frac{3}{(1-x)^3}\Big[1-x^2+2x\ln x\Big],\nonumber \\
F_1^C(x)&=&\frac{2}{(1-x)^4}\Big[2+3x-6x^2+x^3+6x\ln x\Big],\nonumber \\
F_2^C(x)&=&-\frac{3}{2(1-x)^3}\Big[3-4x+x^2+2\ln x\Big].
\end{eqnarray}
These one-loop corrections mainly rely on the bino/wino masses $M_{1,2}$, the Higgsino mass $\mu$, the left- and right-smuon mass parameters,
$M_{\widetilde{\mu}_L,\widetilde{\mu}_R}$, and the ratio of the two Higgs vacuum expectation values, $\tan\beta$. They have a weak dependence on the second generation trilinear coupling $A_\mu$. In the limit of large $\tan\beta$, when all the mass scales are roughly of the same order of $M_{SUSY}$, the contributions Eq.~(\ref{eq:a2-1}) and Eq.~(\ref{eq:a2-2}) can be approximately written as
\begin{eqnarray}
a_{\mu}^{\widetilde\chi^{\pm}}&\simeq&\frac{m_{\mu}^2 g^{2}_2}{32\pi^2 M_{SUSY}^2}\tan\beta; \\
a_{\mu}^{\widetilde\chi^0}&\simeq&\frac{m_{\mu}^2}{192\pi^2 M_{SUSY}^2}\Big(g_1^2-g_2^2\Big)\tan\beta.
\end{eqnarray}
The detailed dependence of $a_\mu$ on the five relevant mass parameters $\tan\beta$ is complicated. For two-loop corrections, it should be noted that if the squark masses (or masses of the first or third generation slepton) become large, the SUSY contributions to $a_\mu$ do not decouple but are logarithmically enhanced. Depending on the mass pattern, a positive or negative correction of O(10\%) for squark masses in the few TeV region can be obtained, see Ref.~\cite{g2-mssm-2loop}.

\section{Constraints on MSSM Explanation of $(g-2)_\mu$} \label{constraints}
In the following, we numerically calculate $\Delta a_{\mu}$ by using the \texttt{FeynHiggs-2.12.0} \cite{feynhiggs} package and scan the relevant MSSM parameter space:
\begin{eqnarray}
10~<\tan\beta<~50, \quad  -2~{\rm TeV} <M_{1}, M_{2} <2~{\rm TeV}, \nonumber \\  -2~{\rm TeV} < \mu < 2~{\rm TeV}, \quad  0.1~{\rm TeV} < m_{\widetilde{l}_L},m_{\widetilde{l}_R} < 2~{\rm TeV.}\label{scan}
\end{eqnarray}
where we have the subscript $\ell=e,\mu$. Due to the small effects on $a_\mu$, the slepton trilinear parameters of the first two generation are assumed as $A_\ell=0$. We also decouple the stau sector by setting the soft stau mass parameters $m_{\tilde{\tau}_L}=m_{\tilde{\tau}_R}=5$ TeV and trilinear parameter $A_{\tau}=0$. So the stau will not contribute to the trilepton signals in our simulations. To satisfy the 125 GeV Higgs mass within a 2 GeV deviation, we vary the stop trilinear parameter in the range $|A_{t}|<5$ TeV and set the stop soft masses at 5 TeV. We require the mixing parameter $|X_{t}/M_{S}|<2$ to avoid the charge/colour-breaking minima \cite{mssm-color-breaking}. We additionally calculate the Higgs mass and the rest of the sparticle masses with \texttt{FeynHiggs-2.12.0} \cite{feynhiggs}.

\subsection{LEP and Higgs data}

In our scan, we also consider the following experimental bounds:
\begin{itemize}
	\item LEP: the direct searches for the slepton and chargino at LEP produce the lower mass limits on the first two generation sleptons and lightest chargino \cite{pdg}:
	\begin{eqnarray}
	m_{\widetilde{l}_L},m_{\widetilde{l}_R} &>& 100\,\text{GeV} \quad (l=e,\mu) \\
	m_{\widetilde{\chi}^{\pm}_1} &>& 105\,\text{GeV}
	\end{eqnarray}
	\item Higgs data: the exclusion limits at 95\% CL from the experimental cross sections from higgs searches at LEP, Tevatron and LHC are examined by using \texttt{HiggsBounds-4.2.1} \cite{higgsbounds}.
    \item We require the lightest neutralino $\widetilde{\chi}^0_1$ as the LSP and $m_{\widetilde{\chi}_1^0}>30\,\text{GeV}$ to be consistent with the bound on light MSSM neutralino dark matter \cite{Calibbi:2014lga}.
\end{itemize}

\begin{figure}[h]
	\centering
	\includegraphics[width=16 cm, height=12 cm]{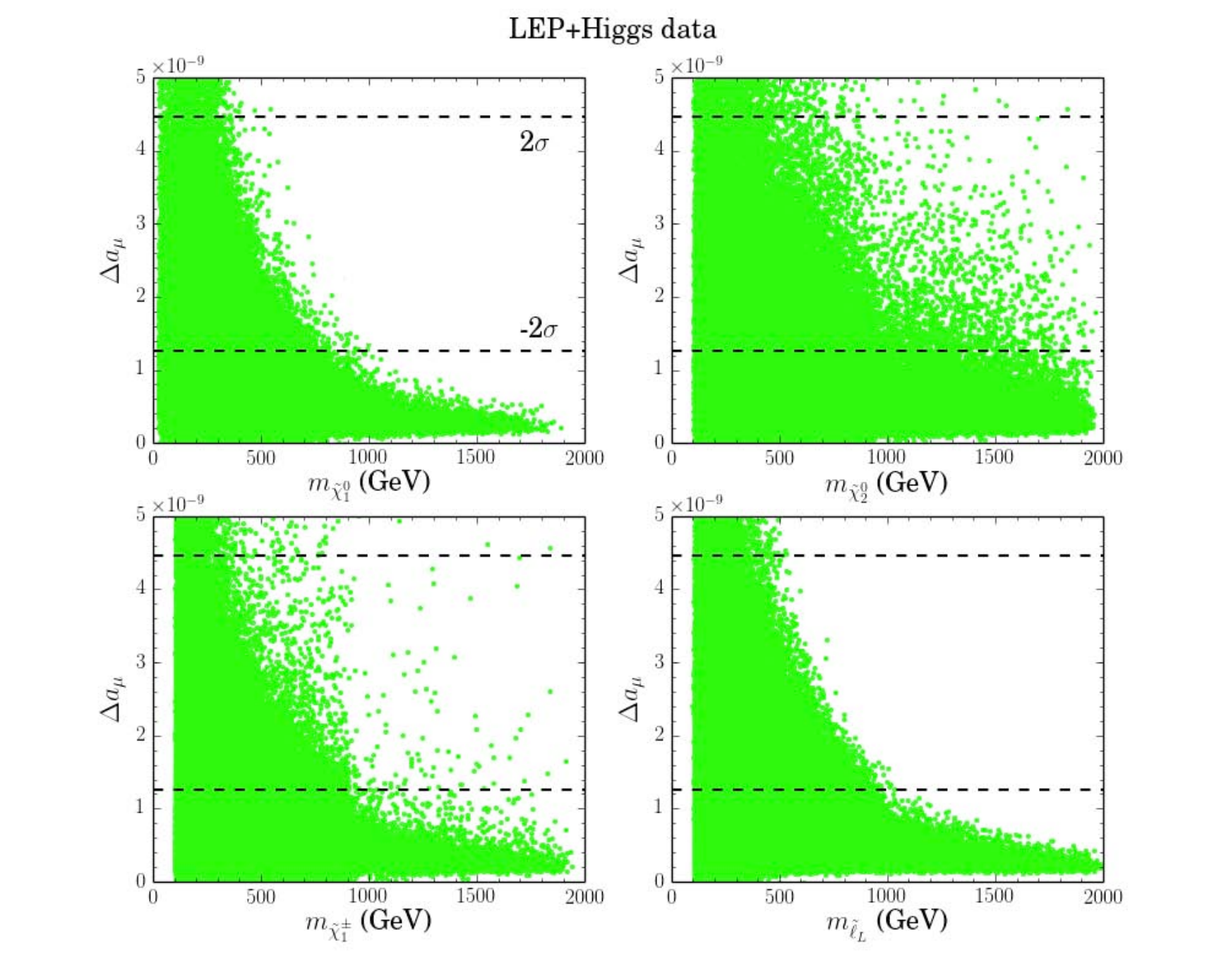}
	\caption{Scatter plot on the plane of $\Delta{a}_{\mu}$ and sparticle masses. Green circles satisfy the constraints from LEP and LHC Higgs data. The dashed lines represent the 2$\sigma$ band on $\Delta{a}_{\mu}$ given by Eq.(\ref{deviation}).}
	\label{g2plots}
\end{figure}
In Fig.~\ref{g2plots}, we present the dependence of $\Delta a_{\mu}$ on the masses of neutralinos ($\widetilde{\chi}^0_{1,2}$), charginos ($\widetilde{\chi}^\pm_{1,2}$) and smuons ($\widetilde{\mu}_{1,2}$). Within the scan ranges of Eq.~\ref{scan}, We find that the $\widetilde{\chi}^\pm$-$\widetilde{\nu}_\mu$ loop dominates over the $\widetilde{\chi}^0$-$\widetilde{\mu}$ loop. A sizable SUSY contribution to $a_\mu$ can be obtained, if $M_1$, $M_2$ and $\mu$ have the same sign and $\widetilde{\chi}^0_{1,2}$ and $\widetilde{\chi}^\pm_{1}$ have a sizable higgsino, wino or both components with large $\tan\beta$. The explanation of $\Delta a_\mu$ within a $2\sigma$ range requires $m_{\widetilde{\chi}^0_{1}} < 1.0$ TeV and $m_{\widetilde{\mu}_{1}} < 1.03$ TeV \footnote{It should be noted that if the higgsino mass parameter $\mu$ is large enough, the $g-2$ anomaly may be explained through the bino-smuon loop contribution, due to the large smuon left-right mixing \cite{g2-Binosmuon}. But such a large $\mu$ scenario is disfavored by the vacuum stability \cite{g2-Binosmuon}, the naturalness \cite{nath} and are highly constrained by the dark matter relic density \cite{ArkaniHamed:2006mb}.}. However, a higgsino or wino-like LSP typically cannot satisfy the constraints of the dark matter relic density and are constrained using data from direct detection experiments.

\subsection{DM relic density and direct detection experiments}

Next, we confront the MSSM explanation of $(g-2)_\mu$ with the various dark matter experiments. We use \texttt{MicrOmegas-4.2.3} \cite{micromega} to calculate the dark matter relic density $\Omega h^2$ and the spin-independent neutralino scattering cross sections with nuclei, denoted as $\sigma^{SI}$.
It should be noted that the thermal relic abundance of the light higgsino or wino-like neutralino dark matter is typically low due to the large annihilation rate in the early universe. This leads to the standard thermally produced WIMP dark matter being under-abundant. In order to have the correct relic density, several alternatives have been proposed, such as choosing the axion-higgsino admixture as a dark matter candidate \cite{axion}. So we rescale the scattering cross section $\sigma^{SI}$ by a factor of ($\Omega h^2/\Omega_{Planck}h^2$), where $\Omega_{Planck}h^2 = 0.112 \pm 0.006$
is the relic density measured by Planck satellite \cite{planck}.

\begin{figure}[t]
	\centering
    \includegraphics[width=0.42\textwidth]{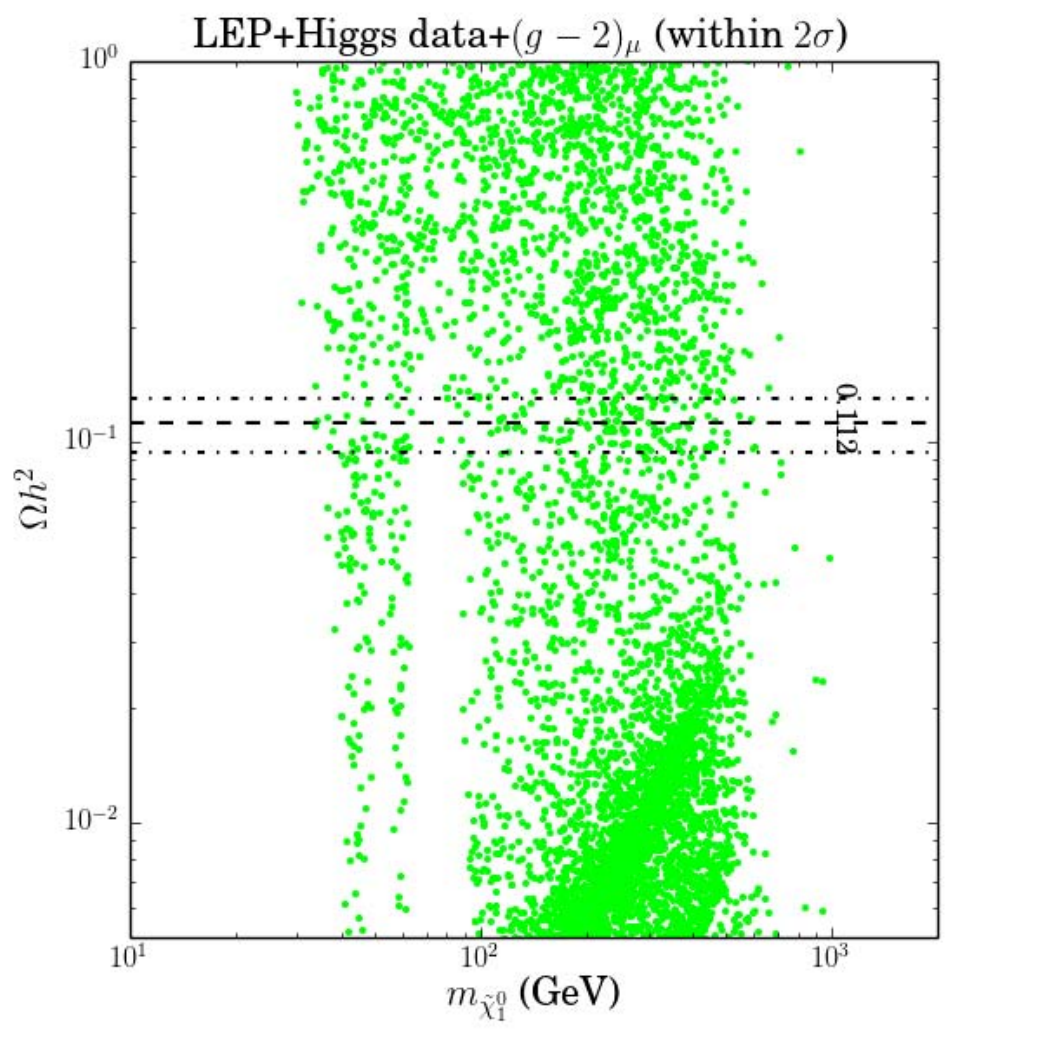}
    \includegraphics[width=0.42\textwidth]{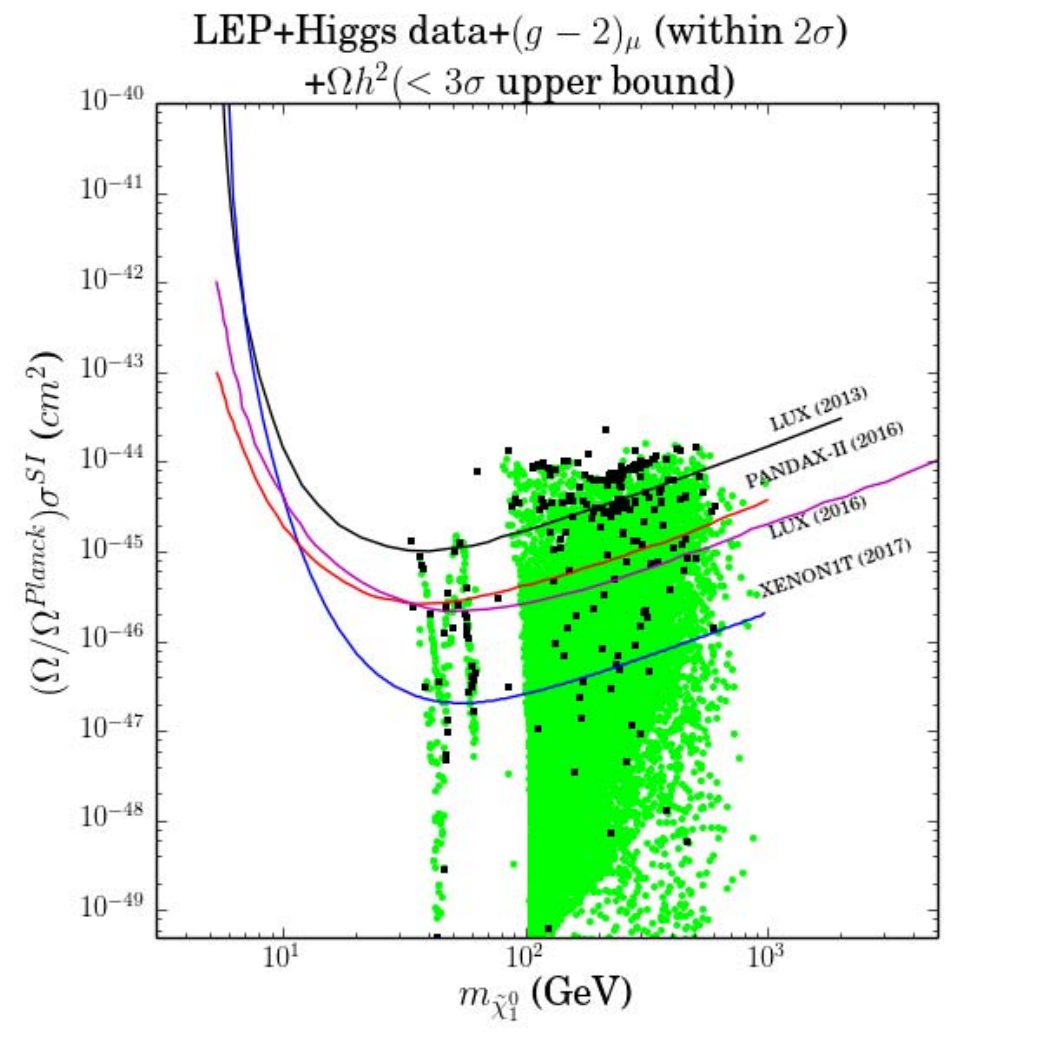}
	\caption{The neutralino dark matter relic density $\Omega h^2$ (left) and the spin-independent neutralino-nucleon scattering cross section $\sigma^{SI}$ (right). The dashed line is the PLANCK central value and the dashed-dotted lines are corresponding $3\sigma$ bands. The exclusion limits on the $\sigma^{SI}$ from LUX (2013) \cite{Akerib:2013tjd} (black line), LUX (2016) (magenta line) \cite{lux}, PandaX-II (red line) \cite{pandaX}, and XENON1T (2017) projected \cite{xenon1t} (blue line). Green circles satisfy the LEP, Higgs data and $2\sigma$ bound of $(g-2)_{\mu}$ (left) and $3\sigma$ upper bound of $\Omega h^2$, while the black squares further require $\Omega h^2$ within $3\sigma$ range. } \label{DMfig}
\end{figure}
In Fig.~\ref{DMfig}, we show the neutralino dark matter relic density $\Omega h^2$ (left) and the spin-independent neutralino-nucleon scattering cross section $\sigma^{SI}$ (right). All samples satisfy the LEP, Higgs data and $(g-2)_{\mu}$ within 2$\sigma$. In the left panel of Fig.~\ref{DMfig}, it can be seen that there are an amount of samples above the $3\sigma$ upper bound of the Planck relic density measurement. Those samples are bino-like and annihilate to the SM particles very slowly, which leads to an overabundance of dark matter in the universe. On the other hand, there are two dips around $M_Z$ and $M_h$, respectively, where $\widetilde{\chi}^0_1\widetilde{\chi}^0_1$ can efficiently annihilate through the resonance effect. When the LSP higgsino or wino component dominates, the annihilation cross section of $\widetilde{\chi}^0_1\widetilde{\chi}^0_1$ is small so that the relic density is less than the $3\sigma$ lower bound of the Planck value. A mixed LSP with a certain higgsino or wino fraction \cite{ArkaniHamed:2006mb} can be reconciled with the measured relic abundance $\Omega h^2$ within the $3\sigma$ range. In the right panel of Fig.~\ref{DMfig}, we project the samples that satisfy $3\sigma$ upper bound $\Omega_{Planck} h^2$ on the plane of $\sigma^{SI}$ versus $m_{\widetilde{\chi}^0_1}$.

A significant portion of the parameter space where the LSP has a sizable higgsino or wino component is excluded by the  recent PandaX-II \cite{pandaX} and LUX data \cite{lux}. The samples with nearly pure higgsino or wino LSPs escape experimental constraints due to the large reduction in the DM abundance. We also find some samples with the correct DM relic density (within 3$\sigma$) and satisfying the LUX constraints. These samples can be placed in two categories. The smaller portion of samples belong to the so called MSSM blind-spot region of parameters \cite{Cheung:2012qy,Han:2016qtc} where the LSP coupling to the Higgs/Z boson is so small that the DM-nucleon scattering cross section is highly suppressed. The sfermions and other heavy higgs bosons are decoupled for these particular samples. The second case is that the bino-like LSPs coannihilate with the sleptons. The scattering cross section of the bino-like LSP with the nucleon can be small to avoid the LUX bound. The future XENON1T (2017) experiment \cite{xenon1t} will further cover the these parameter space.

\begin{figure}[t]
	\centering
	\includegraphics[width=0.5\textwidth]{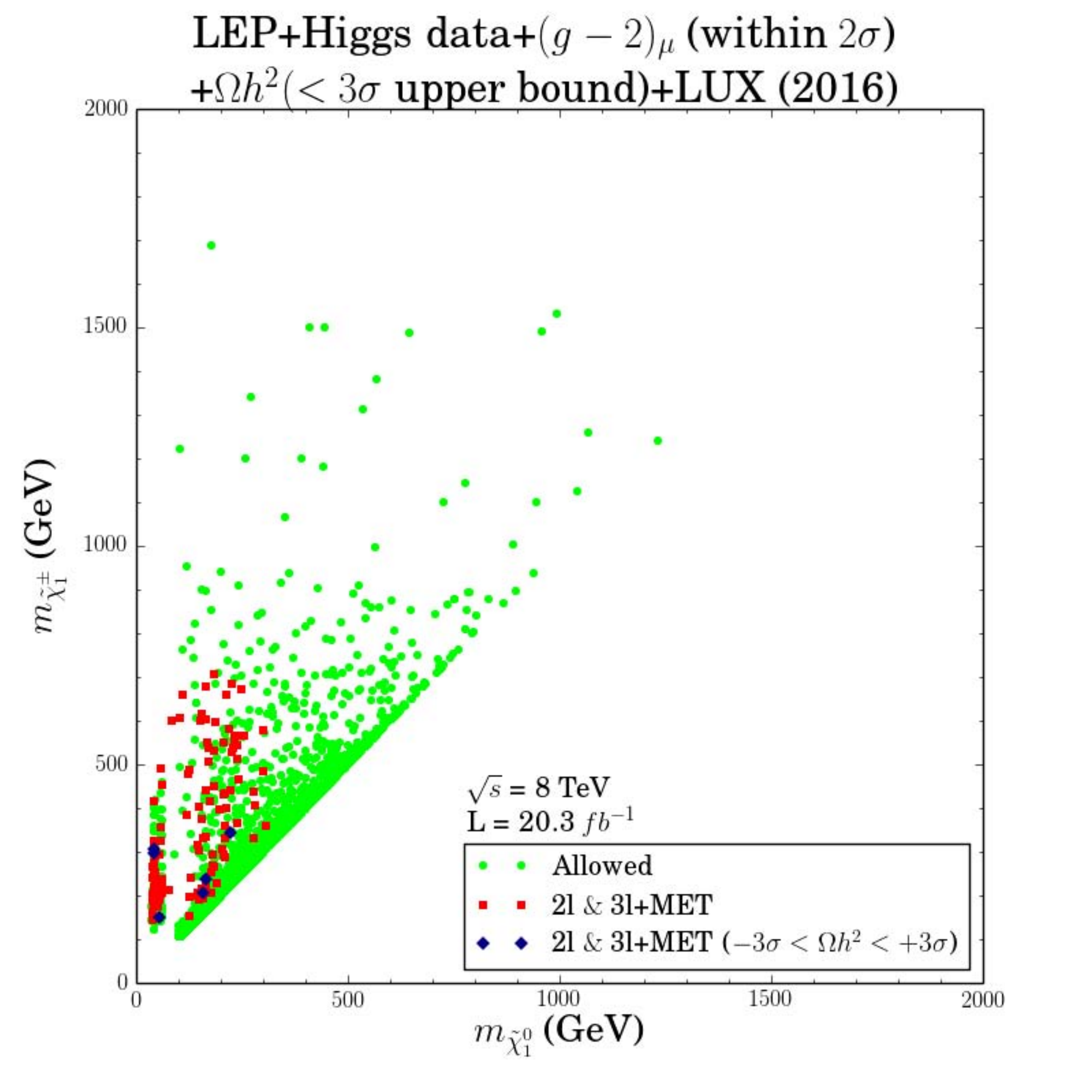}
	\caption{Exclusion limits from LHC Run-1 dilepton and trilepton events. All samples satisfy the LEP, Higgs data, $3\sigma$ upper bound of the dark matter relic density, LUX 2016 and $(g-2)_{\mu}$ within the $2\sigma$. Red squares ($\Omega h^2<+3\sigma$) and blue diamonds ($-3\sigma<\Omega h^2<+3\sigma$) are excluded by $2\ell+\cancel{E}_{T}$ and $3\ell+\cancel{E}_{T}$ events.} \label{fig:8tev}
\end{figure}

\subsection{LHC 8 TeV collider search}

Given the great progress of LHC experiments, we recast the results of searching for $2\ell+\cancel{E}_{T}$ and $3\ell+\cancel{E}_{T}$ signatures at LHC-8 TeV. We focus on 8 TeV data. In fact, most of dedicated analyses at 13 TeV are either preliminary \cite{atlas-13tev-ewkino-1,cms-13tev-ewkino-1,cms-13tev-ewkino-2} or do not provide stronger constraints in general due to the still small luminosity \cite{atlas-13tev-ewkino-2}. The main processes contributing to $2\ell+\cancel{E}_{T}$ events can arise from the production of sleptons pair and charginos:
\begin{equation}
pp\rightarrow\widetilde{\ell}^{+}\widetilde{\ell}^{-},\widetilde{\chi}_{1}^{+}\widetilde{\chi}_{1}^{-}
\end{equation}
with the subsequent decays to leptons:
\begin{itemize}
	\item slepton decay: $\widetilde{\ell}^{\pm}\rightarrow\ell^{\pm}\widetilde{\chi}_{1}^{0}$;
	\item chargino decays: (a) through sleptons: $\widetilde{\chi}_{1}^{\pm}\rightarrow\widetilde{\ell}^{\pm}(\rightarrow\ell^{\pm}\widetilde{\chi}_{1}^{0})\nu_{\ell}$, (b) through sneutrinos: $\widetilde{\chi}_{1}^{\pm}\rightarrow\widetilde{\nu}_{\ell}(\rightarrow\nu_{\ell}\widetilde{\chi}_{1}^{0})\ell^{\pm}$, (c) through $W$ boson: $\widetilde{\chi}_{1}^{\pm}\rightarrow W^\pm(\rightarrow\ell^{\pm}\nu_{\ell}) \widetilde{\chi}^0_1$.
\end{itemize}
While $3\ell+\cancel{E}_{T}$ events mainly come from the associated production of chargino and neutralino:
\begin{equation}
pp\rightarrow\widetilde{\chi}_{i}^{0}\widetilde{\chi}_{j}^{\pm}
\end{equation}
where $i=2,3,4$ and $j=1,2$. They then decay in two different ways:
\begin{itemize}
	\item through sleptons/sneutrinos: (a) $\widetilde{\chi}_{i}^{0}\rightarrow\ell^{\mp}\widetilde{\ell}^{\pm}(\rightarrow\ell^{\pm}\widetilde{\chi}_{1}^{0})$, $\widetilde{\chi}_{j}^{\pm}\rightarrow\widetilde{\ell}^{\pm}(\rightarrow\ell^{\pm}\widetilde{\chi}_{1}^{0})\nu_{\ell}$, (b) $\widetilde{\chi}_{i}^{0}\rightarrow\ell^{\mp}\widetilde{\ell}^{\pm}(\rightarrow\ell^{\pm}\widetilde{\chi}_{1}^{0})$, $\widetilde{\chi}_{j}^{\pm}\rightarrow\widetilde{\nu}_{\ell}(\rightarrow\nu_{\ell}\widetilde{\chi}_{1}^{0})\ell^{\pm}$;
	\item through the SM gauge bosons: $\widetilde{\chi}_{i}^{0}\rightarrow Z^{(*)}(\rightarrow \ell^\pm \ell^\mp)\widetilde{\chi}_{1}^{0}$, $\widetilde{\chi}_{j}^{\pm}\rightarrow W^{\pm(*)}(\rightarrow \ell^\pm \nu_\ell)\widetilde{\chi}_{1}^{0}$.
\end{itemize}
We use \texttt{SPheno-3.3.8} \cite{spheno} to produce the SLHA file to employ in \texttt{MadGraph5\_aMC@NLO} \cite{mad5} and generate the parton level signal events. Then the events are showered and hadronized by \textsf{PYTHIA} \cite{pythia}. The detector effects are included by using the tuned \textsf{Delphes} \cite{delphes}. \texttt{FastJet} \cite{fastjet} is used to cluster jets with the anti-$k_t$ algorithm \cite{anti-kt}. We recast the ATLAS dilepton \cite{lhc-slepton-atlas} and trilepton \cite{lhc-ewkino-atlas} analyses by using \texttt{CheckMATE-1.2.2} \cite{checkmate}. We include the NLO correction effects in the production of $\widetilde{\ell}^\pm \widetilde{\ell}^\mp$, $\widetilde{\chi}^\pm_i \widetilde{\chi}^\mp_i$ and $\widetilde{\chi}^0_i \widetilde{\chi}^\pm_j$ productions by multiplying a $K$-factor $1.3$ \cite{prospino}. The main SM backgrounds include $WZ$, $ZZ$ and $ttV (V=W,Z)$. To estimate the exclusion limit, we define the ratio $r = max(N_{S,i}/S^{95\%}_{obs,i})$, where $N_{S,i}$ and $S^{95\%}_{obs,i}$ are the event numbers of the signal for $i$-th signal region and the corresponding observed 95\% C.L. upper limit, respectively. The max is over all signal regions defined in the analysis. We conclude that a sample is excluded at 95\% C.L., if $r > 1$.

In Fig.~\ref{fig:8tev}, we recast the LHC Run-1 dilepton and trilepton exclusion limits on the plane of $m_{\widetilde{\chi}^\pm_1}$ and $m_{\widetilde{\chi}^0_1}$. All samples satisfy the LEP, Higgs data, $3\sigma$ upper bound of relic density, LUX 2016 and $(g-2)_{\mu}$ within $2\sigma$ range. Red squares ($\Omega h^2<+3\sigma$) and blue diamonds ($-3\sigma<\Omega h^2<+3\sigma$) are excluded by $2\ell+\cancel{E}_{T}$ and $3\ell+\cancel{E}_{T}$ events.  In Fig.~\ref{fig:8tev}, we can see that a portion of samples in $\widetilde{\chi}^\pm_1<710$ GeV and $\widetilde{\chi}^0_1<300$ GeV can be excluded. A bulk of samples in the parameter space with $\widetilde{\chi}^0_1$ being higgsino or wino-like can not be covered because of the small mass difference between $\widetilde{\chi}^\pm_1$ and $\widetilde{\chi}^0_1$. Such a region may be accessed by the monojet(-like) or the VBF production at HL-LHC \cite{lhc-compressed-1,lhc-compressed-2,lhc-compressed-3,lhc-compressed-4,lhc-compressed-4.5,lhc-compressed-5,lhc-compressed-6}. In addition, when $\widetilde{\chi}^0_2$ has a sizable bino component, the limit from trilepton events will become weak because of the reduction of cross section of $\widetilde{\chi}^\pm_1\widetilde{\chi}^0_2$. We also find that the dilepton channel can be complimentary to the trilepton channel when the latter is suppressed by small neutralino leptonic branching ratios. An important factor in the dilepton and trilepton yields is the leptonic branching fraction which can vary widely throughout the parameter space. If the slepton is on shell, the chargino two-body decays then dominate and its leptonic branching fraction is maximized, $Br(\widetilde{\chi}^\pm_1 \to \widetilde{\chi}^0_1 \widetilde{\ell}^\pm(\rightarrow \ell^\pm \nu_\ell))_{max}=2/3$. When the sneutrino is on-shell and is lighter than the corresponding slepton, the channel $\widetilde{\chi}^0_2 \to \nu_\ell \widetilde{\nu}_\ell$ will dominate the decay width, and the neutralino leptonic branching ratio is suppressed. On the other hand, if the slepton and sneutrino are heavy enough, the decay amplitudes of $\widetilde{\chi}^\pm_1$ and $\widetilde{\chi}^0_2$ are dominated by $W$ and $Z$ boson exchange, respectively, which give $\widetilde{\chi}^\pm_1 \to \widetilde{\chi}^0_1 W^\pm(\to \ell^\pm \nu_\ell) \simeq 2/9$ and $\widetilde{\chi}^0_2 \to \widetilde{\chi}^0_1 Z(\to \ell^\pm \ell^\mp) \simeq 6\%$. On the other hand, $\widetilde{\chi}^0_2$ can decay to $h\widetilde{\chi}^0_1$ with a sizable branching ratio if kinematically accessible, which can also weaken the trilepton exclusion limit.

\section{Prospects at a 100 TeV Collider} \label{100tev}
To hunt for new fundamental particles, a 100 TeV $pp$ collider has been under discussion in recent years, which will allow us to probe the new physics scale roughly an order of magnitude higher than we can possibly reach with the LHC \cite{Arkani-Hamed:2015vfh}. In this section, we estimate the prospects of probing the MSSM explanation of the $(g-2)_{\mu}$ anomaly by extrapolating the above 8 TeV trilepton analysis to a 100 TeV $pp$ collider. For each allowed sample above, we use the most sensitive signal region in 8 TeV analysis and simply assume the same detection efficiency in the 100 TeV analysis. We rescale the signal ($S$) and background ($B$) events by the following ratio:
\begin{equation}
N^{100\,\text{TeV}}=(\sigma^{100\,\text{TeV}}/\sigma^{8\,\text{TeV}})(3000\,\text{fb}^{-1}/20.3\,\text{fb}^{-1})N^{8\,\text{TeV}}
\end{equation}
Such a treatment can be considered as a preliminary theoretical estimation. The optimized analysis strategy may be achieved once the details of the collider environment is known. To obtain the expected exclusion limits, we use the following equation,
\begin{equation}
\frac{S}{\sqrt{B+(\beta_{sys} B)^2}} \geq 2 \quad [\text{Excluded}] \label{excluded}
\end{equation}
where the factor $\beta_{sys}$ parameterizes the systematic uncertainty. In Fig.~\ref{100tev}, we can see that when $\beta_{sys}=0.1$, a majority of samples allowed by $(g-2)_\mu$ in the parameter space with $\widetilde{\chi}^0_1<530$ GeV and $\widetilde{\chi}^\pm_1<940$ GeV can be excluded. Such a range will be extended to $\widetilde{\chi}^0_1<710$ GeV and $\widetilde{\chi}^\pm_1<940$ GeV, if $\beta_{sys}=0$.

It should be noted that the region that satisfies the DM relic density within the $3\sigma$ range through the $Z$ or $h$ resonant annihilation in the blind spots can be covered by searching for trilepton events from $\tilde{\chi}^0_2\tilde{\chi}^+_1$ associated production at a 100 TeV $pp$ collider. The samples that are beyond future sensitivity of this trilepton search and the DM direct detections are either higgsino/wino-like LSPs with the compressed mass spectrum or bino-like LSPs co-annihilating with sleptons. Such compressed regions may be probed by the monojet(-like) searches at a 100 TeV $pp$ collider \cite{100tev-ewkino-4}.

\begin{figure}[t]
	\centering
	\includegraphics[width=\textwidth]{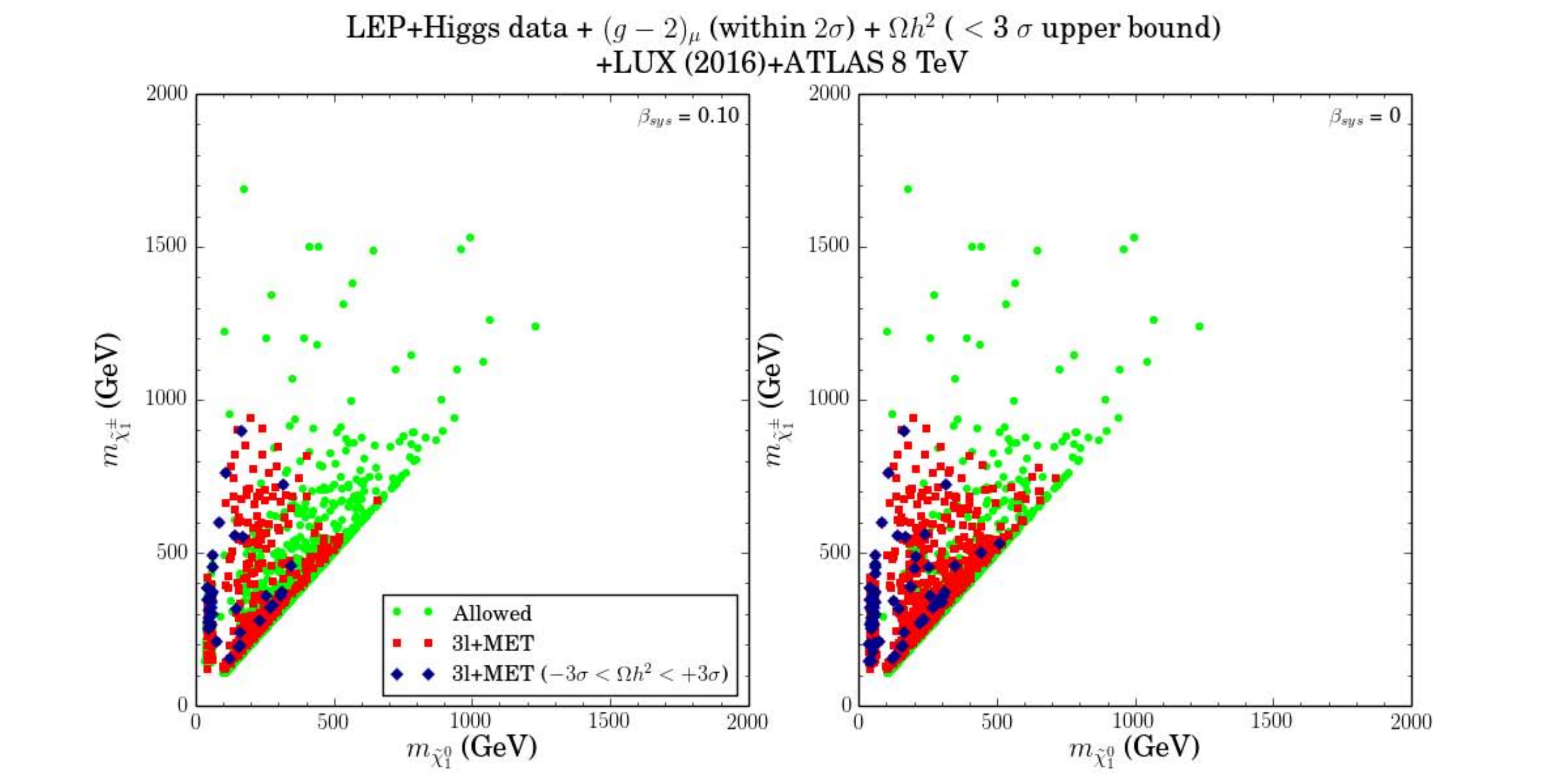}
	\caption{Same as Fig.~\ref{fig:8tev}, but for expected exclusion limit at a 100 TeV $pp$ collider with the luminosity of 3000 fb$^{-1}$. Red squares ($\Omega h^2<+3\sigma$) and blue diamonds $-3\sigma<\Omega h^2<+3\sigma$ are excluded by searching for $3\ell$ + MET events. The systematic uncertainty $\beta_{sys}$ is taken as 0.1 and 0, respectively.} \label{fig:100tev}
\end{figure}

\section{Conclusion} \label{concl}
In this work we have studied the prospect of current and future dark matter and collider experiments in probing the anomalous magnetic moment of the muon in the MSSM. Under the constraints of Higgs data, dark matter relic density, PandaX-II/LUX-2016 experiments and LHC-8 TeV searches for dilepton/trilepton events, we find the Planck data and the recent PandaX-II/LUX data can significantly exclude the MSSM parameter space satisfying $(g-2)_{\mu}$, which will be further excluded by the upcoming XENON-1T (2017) experiment. We also find that most of our surviving samples that satisfy DM relic density within $3\sigma$ range through $Z$ or $h$ resonant effect can be covered by searching for trilepton events from $\tilde{\chi}^0_2\tilde{\chi}^+_1$ associated production a 100 TeV $pp$ collider. While the samples that are beyond the future sensitivity of this trilepton search and DM direct detections are either higgsino/wino-like LSPs or bino-like LSPs co-annihilating with sleptons. Such compressed regions may be probed by the monojet(-like) searches at a future 100 TeV $pp$ collider.

\paragraph{Acknowledgement.}
This work was partially supported by the Australian Research Council. LW was also supported in part by the National Natural Science Foundation of China (NNSFC) under grants Nos. 11305049, 11275057.


\begin{thebibliography}{999}
	
\bibitem{higgs-atlas}
  G.~Aad {\it et al.}  [ATLAS Collaboration],
  %``Observation of a new particle in the search for the Standard Model Higgs boson with the ATLAS detector at the LHC,''
  Phys.\ Lett.\ B {\bf 716}, 1 (2012).
  %%CITATION = ARXIV:1207.7214;%%
  %1339 citations counted in INSPIRE as of 12 Jul 2013kp

\bibitem{higgs-cms}
  S.~Chatrchyan {\it et al.}  [CMS Collaboration],
  %``Observation of a new boson at a mass of 125 GeV with the CMS experiment at the LHC,''
  Phys.\ Lett.\ B {\bf 716}, 30 (2012).
  %%CITATION = ARXIV:1207.7235;%%
  %1322 citations counted in INSPIRE as of 12 Jul 2013

%\cite{Hagiwara:2002ma}
\bibitem{g2-review-1}
  K.~Hagiwara, A.~D.~Martin, D.~Nomura and T.~Teubner,
  %``The SM prediction of g-2 of the muon,''
  Phys.\ Lett.\ B {\bf 557}, 69 (2003)
  doi:10.1016/S0370-2693(03)00138-2
  [hep-ph/0209187].
  %%CITATION = doi:10.1016/S0370-2693(03)00138-2;%%
  %172 citations counted in INSPIRE as of 01 Aug 2016

\bibitem{g2-review-2}
  F.~Jegerlehner and A.~Nyffeler,
  %``The Muon g-2,''
  Phys.\ Rept.\ {\bf 477}, (2009) 1.
  %%CITATION = PRPLC,477,1;%%

\bibitem{g2-review-3}
  J.~P.~Miller, E.~de Rafael, B.~L.~Roberts and D.~St\"ockinger,
  %``Muon (g-2): Experiment and Theory,''
  Ann.\ Rev.\ Nucl.\ Part.\ Sci.\ {\bf 62}, 237 (2012).
  %%CITATION = ARNUA,62,237;%%

\bibitem{g2-review-4}
  T.~Blum, A.~Denig, I.~Logashenko, E.~de Rafael, B.~L.~Roberts, T.~Teubner and G.~Venanzoni,
  %``The Muon (g-2) Theory Value: Present and Future,''
  arXiv:1311.2198 [hep-ph].
  %%CITATION = ARXIV:1311.2198;%%



%\cite{Bennett:2006fi}
\bibitem{g2-exp}
  G.~W.~Bennett {\it et al.} [Muon g-2 Collaboration],
  %``Final Report of the Muon E821 Anomalous Magnetic Moment Measurement at BNL,''
  Phys.\ Rev.\ D {\bf 73}, 072003 (2006).
  %%CITATION = doi:10.1103/PhysRevD.73.072003;%%
  %1218 citations counted in INSPIRE as of 01 Aug 2016


\bibitem{g2-sm-1}
%\cite{Davier:2010nc}
  M.~Davier, A.~Hoecker, B.~Malaescu and Z.~Zhang,
  %``Reevaluation of the Hadronic Contributions to the Muon g-2 and to alpha(MZ),''
  Eur.\ Phys.\ J.\ C {\bf 71}, 1515 (2011)
  Erratum: [Eur.\ Phys.\ J.\ C {\bf 72}, 1874 (2012)].
  %%CITATION = doi:10.1140/epjc/s10052-012-1874-8, 10.1140/epjc/s10052-010-1515-z;%%
  %537 citations counted in INSPIRE as of 01 Aug 2016


%\cite{Hagiwara:2011af}
\bibitem{g2-sm-2}
  K.~Hagiwara, R.~Liao, A.~D.~Martin, D.~Nomura and T.~Teubner,
  %``(g-2)_mu and alpha(M_Z^2) re-evaluated using new precise data,''
  J.\ Phys.\ G {\bf 38}, 085003 (2011).
  %%CITATION = doi:10.1088/0954-3899/38/8/085003;%%
  %427 citations counted in INSPIRE as of 01 Aug 2016


%\cite{Aoyama:2012wk}
\bibitem{g2-sm-qed}
  T.~Aoyama, M.~Hayakawa, T.~Kinoshita and M.~Nio,
  %``Complete Tenth-Order QED Contribution to the Muon g-2,''
  Phys.\ Rev.\ Lett.\  {\bf 109}, 111808 (2012).
  %%CITATION = doi:10.1103/PhysRevLett.109.111808;%%
  %167 citations counted in INSPIRE as of 01 Aug 2016

%\cite{Gnendiger:2013pva}
\bibitem{g2-sm-ew}
  C.~Gnendiger, D.~Stöckinger and H.~Stöckinger-Kim,
  %``The electroweak contributions to $(g-2)_\mu$ after the Higgs boson mass measurement,''
  Phys.\ Rev.\ D {\bf 88}, 053005 (2013).
  %%CITATION = doi:10.1103/PhysRevD.88.053005;%%
  %77 citations counted in INSPIRE as of 01 Aug 2016


\bibitem{g2-exp-new}
  G.~Venanzoni [Fermilab E989 Collaboration],
  %``The New Muon g−2 experiment at Fermilab,''
  Nucl.\ Part.\ Phys.\ Proc.\  {\bf 273-275}, 584 (2016).
  %%CITATION = doi:10.1016/j.nuclphysbps.2015.09.087;%%
  %12 citations counted in INSPIRE as of 01 Aug 2016

%\cite{Moroi:1995yh}
\bibitem{g2-mssm-0}
  T.~Moroi,
  %``The Muon anomalous magnetic dipole moment in the minimal supersymmetric standard model,''
  Phys.\ Rev.\ D {\bf 53}, 6565 (1996)
  Erratum: [Phys.\ Rev.\ D {\bf 56}, 4424 (1997)].
  %%CITATION = doi:10.1103/PhysRevD.53.6565, 10.1103/PhysRevD.56.4424;%%
  %453 citations counted in INSPIRE as of 01 Aug 2016


\bibitem{g2-mssm-2}
  S.~P.~Martin and J.~D.~Wells,
  %``Muon anomalous magnetic dipole moment in supersymmetric theories,''
  Phys.\ Rev.\ D {\bf 64}, 035003 (2001).
  %%CITATION = doi:10.1103/PhysRevD.64.035003;%%
  %237 citations counted in INSPIRE as of 01 Aug 2016

\bibitem{g2-mssm-3}
D.~St\"ockinger,
%``The muon magnetic moment and
%supersymmetry,''
J.\ Phys.\ G {\bf 34}, R45 (2007).
 % [arXiv:hep-ph/0609168].
%%CITATION = JPHGB,G34,R45;%%

 %\cite{Cho:2011rk}
\bibitem{g2-mssm-6}
  G.~-C.~Cho, K.~Hagiwara, Y.~Matsumoto and D.~Nomura,
  %``The MSSM confronts the precision electroweak data and the muon g-2,''
  JHEP {\bf 1111}, 068 (2011).
  %%CITATION = ARXIV:1104.1769;%%
  %31 citations counted in INSPIRE as of 07 Aug 2013


\bibitem{g2-mssm-10}
  M.~Endo, K.~Hamaguchi, S.~Iwamoto and T.~Yoshinaga,
  %``Muon g-2 vs LHC in Supersymmetric Models,''
  JHEP {\bf 1401}, 123 (2014).
  %%CITATION = ARXIV:1303.4256;%%
  %45 citations counted in INSPIRE as of 26 Sep 2015
%
%

%\cite{Ibe:2013oha}
\bibitem{g2-mssm-15}
  M.~Ibe, T.~T.~Yanagida and N.~Yokozaki,
  %``Muon g-2 and 125 GeV Higgs in Split-Family Supersymmetry,''
  JHEP {\bf 1308}, 067 (2013).
   %%CITATION = ARXIV:1303.6995;%%

%\cite{Akula:2013ioa}
\bibitem{g2-mssm-16}
  S.~Akula and P.~Nath,
  %``Gluino-driven radiative breaking, Higgs boson mass, muon g-2, and the Higgs diphoton decay in supergravity unification,''
  Phys.\ Rev.\ D {\bf 87}, no. 11, 115022 (2013).
  %%CITATION = doi:10.1103/PhysRevD.87.115022;%%
  %40 citations counted in INSPIRE as of 01 Aug 2016

\bibitem{g2-mssm-17}
  S.~Mohanty, S.~Rao and D.~P.~Roy,
  %``Reconciling the muon $g-2$ and dark matter relic density with the LHC results in nonuniversal gaugino mass models,''
  JHEP {\bf 1309}, 027 (2013).
  %%CITATION = ARXIV:1303.5830;%%



\bibitem{g2-mssm-19}
  K.~Kowalska, L.~Roszkowski, E.~M.~Sessolo and A.~J.~Williams,
  %``GUT-inspired SUSY and the muon g-2 anomaly: prospects for LHC 14 TeV,''
  arXiv:1503.08219 [hep-ph].
  %%CITATION = ARXIV:1503.08219;%%

%\cite{Wang:2015nra}
\bibitem{g2-mssm-20}
  F.~Wang, W.~Wang, J.~M.~Yang and Y.~Zhang,
  %``Heavy colored SUSY partners from deflected anomaly mediation,''
  JHEP {\bf 1507}, 138 (2015).
  %%CITATION = doi:10.1007/JHEP07(2015)138;%%
  %8 citations counted in INSPIRE as of 01 Aug 2016


%\cite{Wang:2015rli}
\bibitem{g2-mssm-21}
  F.~Wang, W.~Wang and J.~M.~Yang,
  %``Reconcile muon g-2 anomaly with LHC data in SUGRA with generalized gravity mediation,''
  JHEP {\bf 1506}, 079 (2015).
  %%CITATION = doi:10.1007/JHEP06(2015)079;%%
  %8 citations counted in INSPIRE as of 01 Aug 2016


%\cite{Wang:2015kuj}
\bibitem{g2-mssm-22}
  F.~Wang, L.~Wu, J.~M.~Yang and M.~Zhang,
  %``750 GeV diphoton resonance, 125 GeV Higgs and muon g − 2 anomaly in deflected anomaly mediation SUSY breaking scenarios,''
  Phys.\ Lett.\ B {\bf 759}, 191 (2016).
  %%CITATION = doi:10.1016/j.physletb.2016.05.071;%%
  %124 citations counted in INSPIRE as of 01 Aug 2016


%\cite{Okada:2013ija}
\bibitem{g2-mssm-23}
  N.~Okada, S.~Raza and Q.~Shafi,
  %``Particle Spectroscopy of Supersymmetric SU(5) in Light of 125 GeV Higgs and Muon g-2 Data,''
  Phys.\ Rev.\ D {\bf 90}, no. 1, 015020 (2014)
  [arXiv:1307.0461 [hep-ph]].
  %%CITATION = ARXIV:1307.0461;%%


\bibitem{g2-mssm-24}
  I.~Gogoladze, F.~Nasir, Q.~Shafi and C.~S.~Un,
  %``Nonuniversal Gaugino Masses and Muon g-2,''
  Phys.\ Rev.\ D {\bf 90} (2014) 3, 035008.
  %%CITATION = ARXIV:1403.2337;%%

%\cite{Babu:2014lwa}
\bibitem{g2-mssm-25}
  K.~S.~Babu, I.~Gogoladze, Q.~Shafi and C.~S.~Ün,
  %``Muon g-2, 125 GeV Higgs boson, and neutralino dark matter in a flavor symmetry-based MSSM,''
  Phys.\ Rev.\ D {\bf 90}, no. 11, 116002 (2014).
  %%CITATION = doi:10.1103/PhysRevD.90.116002;%%
  %11 citations counted in INSPIRE as of 01 Aug 2016

%\cite{Ajaib:2014ana}
\bibitem{g2-mssm-26}
  M.~A.~Ajaib, I.~Gogoladze, Q.~Shafi and C.~S.~Ün,
  %``Split sfermion families, Yukawa unification and muon $g - 2$,''
  JHEP {\bf 1405}, 079 (2014).
  %%CITATION = doi:10.1007/JHEP05(2014)079;%%
  %21 citations counted in INSPIRE as of 01 Aug 2016

\bibitem{g2-mssm-27}
  M.~A.~Ajaib, B.~Dutta, T.~Ghosh, I.~Gogoladze and Q.~Shafi,
  %``Neutralinos and sleptons at the LHC in light of muon $(g-2)_{\mu}$,''
  Phys.\ Rev.\ D {\bf 92}, no. 7, 075033 (2015).
  %%CITATION = doi:10.1103/PhysRevD.92.075033;%%
  %9 citations counted in INSPIRE as of 01 Aug 2016

%\cite{Ajaib:2015ika}
\bibitem{g2-mssm-28}
  M.~Adeel Ajaib, I.~Gogoladze and Q.~Shafi,
  %``GUT-inspired supersymmetric model for h→γγ and the muon g-2,''
  Phys.\ Rev.\ D {\bf 91}, no. 9, 095005 (2015).
  %%CITATION = doi:10.1103/PhysRevD.91.095005;%%
  %9 citations counted in INSPIRE as of 01 Aug 2016

%\cite{Gogoladze:2015jua}
\bibitem{g2-mssm-29}
  I.~Gogoladze, Q.~Shafi and C.~S.~Ün,
  %``Reconciling the muon g−2 , a 125 GeV Higgs boson, and dark matter in gauge mediation models,''
  Phys.\ Rev.\ D {\bf 92}, no. 11, 115014 (2015).
  %%CITATION = doi:10.1103/PhysRevD.92.115014;%%
  %2 citations counted in INSPIRE as of 01 Aug 2016

\bibitem{g2-mssm-30}
  M.~Badziak, Z.~Lalak, M.~Lewicki, M.~Olechowski and S.~Pokorski,
  %``Upper bounds on sparticle masses from muon g - 2 and the Higgs mass and the complementarity of future colliders,''
  JHEP {\bf 1503} (2015) 003.
  %%CITATION = ARXIV:1411.1450;%%

%\cite{Athron:2015rva}
\bibitem{g2-mssm-31}
  P.~Athron {\it et al.},
  %``GM2Calc: Precise MSSM prediction for $(g - 2)$ of the muon,''
  Eur.\ Phys.\ J.\ C {\bf 76}, no. 2, 62 (2016).
  %%CITATION = doi:10.1140/epjc/s10052-015-3870-2;%%
  %4 citations counted in INSPIRE as of 01 Aug 2016


\bibitem{lhc-gluino-atlas}
%\cite{Aad:2013wta}
  G.~Aad {\it et al.} [ATLAS Collaboration],
  %``Search for new phenomena in final states with large jet multiplicities and missing transverse momentum at $\sqrt{s}$=8 TeV proton-proton collisions using the ATLAS experiment,''
  JHEP {\bf 1310}, 130 (2013)
  Erratum: [JHEP {\bf 1401}, 109 (2014)].
  %%CITATION = doi:10.1007/JHEP10(2013)130, 10.1007/JHEP01(2014)109;%%
  %157 citations counted in INSPIRE as of 01 Aug 2016


\bibitem{lhc-gluino-cms}
%\cite{Chatrchyan:2014lfa}
  S.~Chatrchyan {\it et al.} [CMS Collaboration],
  %``Search for new physics in the multijet and missing transverse momentum final state in proton-proton collisions at $\sqrt{s}$= 8 TeV,''
  JHEP {\bf 1406}, 055 (2014).
  %%CITATION = doi:10.1007/JHEP06(2014)055;%%
  %220 citations counted in INSPIRE as of 01 Aug 2016


\bibitem{lhc-stop-atlas}
%\cite{Aad:2014kra}
  G.~Aad {\it et al.} [ATLAS Collaboration],
  %``Search for top squark pair production in final states with one isolated lepton, jets, and missing transverse momentum in $\sqrt s =$8 TeV $pp$ collisions with the ATLAS detector,''
  JHEP {\bf 1411}, 118 (2014).
  %%CITATION = doi:10.1007/JHEP11(2014)118;%%
  %174 citations counted in INSPIRE as of 01 Aug 2016

\bibitem{lhc-stop-cms}
%\cite{Chatrchyan:2013xna}
  S.~Chatrchyan {\it et al.} [CMS Collaboration],
  %``Search for top-squark pair production in the single-lepton final state in pp collisions at $\sqrt{s}$ = 8 TeV,''
  Eur.\ Phys.\ J.\ C {\bf 73}, no. 12, 2677 (2013).
  %%CITATION = doi:10.1140/epjc/s10052-013-2677-2;%%
  %265 citations counted in INSPIRE as of 01 Aug 2016

\bibitem{lhc-steath}
%\cite{Fan:2015mxp}
  J.~Fan, R.~Krall, D.~Pinner, M.~Reece and J.~T.~Ruderman,
  %``Stealth Supersymmetry Simplified,''
  JHEP {\bf 1607}, 016 (2016).
  %%CITATION = doi:10.1007/JHEP07(2016)016;%%
  %4 citations counted in INSPIRE as of 04 Aug 2016

\bibitem{lhc-ns-1}
%\cite{Han:2013kga}
  C.~Han, K.~i.~Hikasa, L.~Wu, J.~M.~Yang and Y.~Zhang,
  %``Current experimental bounds on stop mass in natural SUSY,''
  JHEP {\bf 1310}, 216 (2013).
  %%CITATION = doi:10.1007/JHEP10(2013)216;%%
  %52 citations counted in INSPIRE as of 01 Aug 2016

\bibitem{lhc-ns-2}
%\cite{Kobakhidze:2015scd}
  A.~Kobakhidze, N.~Liu, L.~Wu, J.~M.~Yang and M.~Zhang,
  %``Closing up a light stop window in natural SUSY at LHC,''
  Phys.\ Lett.\ B {\bf 755}, 76 (2016).
  %%CITATION = doi:10.1016/j.physletb.2016.02.003;%%
  %8 citations counted in INSPIRE as of 01 Aug 2016

%\cite{Drees:2015aeo}
\bibitem{lhc-ns-3}
  M.~Drees and J.~S.~Kim,
  %``Minimal natural supersymmetry after the LHC8,''
  Phys.\ Rev.\ D {\bf 93}, no. 9, 095005 (2016).
  %%CITATION = doi:10.1103/PhysRevD.93.095005;%%
  %9 citations counted in INSPIRE as of 04 Aug 2016
  
%\cite{Han:2016gvr}
\bibitem{lhc-ns-4}
  C.~Han, K.~i.~Hikasa, L.~Wu, J.~M.~Yang and Y.~Zhang,
  %``Status of CMSSM in light of current LHC Run-2 and LUX data,''
  arXiv:1612.02296 [hep-ph].
  %%CITATION = ARXIV:1612.02296;%%
  
%\cite{Duan:2016vpp}
\bibitem{lhc-ns-5}
  G.~H.~Duan, K.~i.~Hikasa, L.~Wu, J.~M.~Yang and M.~Zhang,
  %``Leptonic mono-top from single stop production at LHC,''
  arXiv:1611.05211 [hep-ph].
  %%CITATION = ARXIV:1611.05211;%%
  %2 citations counted in INSPIRE as of 07 Mar 2017
  
%\cite{Han:2016xet}
\bibitem{lhc-ns-6}
  C.~Han, J.~Ren, L.~Wu, J.~M.~Yang and M.~Zhang,
  %``Top-squark in natural SUSY under current LHC run-2 data,''
  Eur.\ Phys.\ J.\ C {\bf 77}, no. 2, 93 (2017)
  doi:10.1140/epjc/s10052-017-4662-7
  [arXiv:1609.02361 [hep-ph]].
  %%CITATION = doi:10.1140/epjc/s10052-017-4662-7;%%
  %8 citations counted in INSPIRE as of 07 Mar 2017      

\bibitem{lhc-slepton-atlas}
%\cite{TheATLAScollaboration:2013hha}
  The ATLAS collaboration [ATLAS Collaboration],
  %``Search for direct-slepton and direct-chargino production in final states with two opposite-sign leptons, missing transverse momentum and no jets in 20/fb of pp collisions at sqrt(s) = 8 TeV with the ATLAS detector,''
  ATLAS-CONF-2013-049.
  %%CITATION = ATLAS-CONF-2013-049;%%
  %100 citations counted in INSPIRE as of 01 Aug 2016


\bibitem{lhc-slepton-cms}
%\cite{Khachatryan:2014qwa}
  V.~Khachatryan {\it et al.} [CMS Collaboration],
  %``Searches for electroweak production of charginos, neutralinos, and sleptons decaying to leptons and W, Z, and Higgs bosons in pp collisions at 8 TeV,''
  Eur.\ Phys.\ J.\ C {\bf 74}, no. 9, 3036 (2014).
  %%CITATION = doi:10.1140/epjc/s10052-014-3036-7;%%
  %181 citations counted in INSPIRE as of 01 Aug 2016

\bibitem{lhc-ewkino-atlas}
%\cite{Aad:2014nua}
  G.~Aad {\it et al.} [ATLAS Collaboration],
  %``Search for direct production of charginos and neutralinos in events with three leptons and missing transverse momentum in $\sqrt{s} =$ 8TeV $pp$ collisions with the ATLAS detector,''
  JHEP {\bf 1404}, 169 (2014).
  %%CITATION = doi:10.1007/JHEP04(2014)169;%%
  %168 citations counted in INSPIRE as of 01 Aug 2016


%\cite{Khachatryan:2014mma}
\bibitem{lhc-ewkino-cms}
  V.~Khachatryan {\it et al.} [CMS Collaboration],
  %``Searches for electroweak neutralino and chargino production in channels with Higgs, Z, and W bosons in pp collisions at 8 TeV,''
  Phys.\ Rev.\ D {\bf 90}, no. 9, 092007 (2014).
  %%CITATION = doi:10.1103/PhysRevD.90.092007;%%
  %71 citations counted in INSPIRE as of 01 Aug 2016



%\cite{Ellis:2012aa}
\bibitem{mssm-dm-1}
  J.~Ellis and K.~A.~Olive,
  %``Revisiting the Higgs Mass and Dark Matter in the CMSSM,''
  Eur.\ Phys.\ J.\ C {\bf 72}, 2005 (2012).
  %%CITATION = doi:10.1140/epjc/s10052-012-2005-2;%%
  %142 citations counted in INSPIRE as of 01 Aug 2016

%\cite{Baer:2012uya}
\bibitem{mssm-dm-2}
  H.~Baer, V.~Barger and A.~Mustafayev,
  %``Neutralino dark matter in mSUGRA/CMSSM with a 125 GeV light Higgs scalar,''
  JHEP {\bf 1205}, 091 (2012).
  %%CITATION = doi:10.1007/JHEP05(2012)091;%%
  %120 citations counted in INSPIRE as of 01 Aug 2016

%\cite{Belanger:2013pna}
\bibitem{mssm-dm-3}
  G.~Bélanger, G.~Drieu La Rochelle, B.~Dumont, R.~M.~Godbole, S.~Kraml and S.~Kulkarni,
  %``LHC constraints on light neutralino dark matter in the MSSM,''
  Phys.\ Lett.\ B {\bf 726}, 773 (2013).
  %%CITATION = doi:10.1016/j.physletb.2013.09.059;%%
  %26 citations counted in INSPIRE as of 01 Aug 2016


%\cite{Cahill-Rowley:2014boa}
\bibitem{mssm-dm-4}
  M.~Cahill-Rowley, R.~Cotta, A.~Drlica-Wagner, S.~Funk, J.~Hewett, A.~Ismail, T.~Rizzo and M.~Wood,
  %``Complementarity of dark matter searches in the phenomenological MSSM,''
  Phys.\ Rev.\ D {\bf 91}, no. 5, 055011 (2015).
  %%CITATION = doi:10.1103/PhysRevD.91.055011;%%
  %26 citations counted in INSPIRE as of 01 Aug 2016

%\cite{Fargnoli:2013zia}
\bibitem{g2-mssm-2loop}
  H.~Fargnoli, C.~Gnendiger, S.~Paßehr, D.~Stöckinger and H.~Stöckinger-Kim,
  %``Two-loop corrections to the muon magnetic moment from fermion/sfermion loops in the MSSM: detailed results,''
  JHEP {\bf 1402}, 070 (2014).
  %%CITATION = doi:10.1007/JHEP02(2014)070;%%
  %20 citations counted in INSPIRE as of 01 Aug 2016

\bibitem{feynhiggs}
  S.~Heinemeyer, W.~Hollik and G.~Weiglein,
  %``FeynHiggs: A Program for the calculation of the masses of the neutral CP even Higgs bosons in the MSSM,''
  Comput.\ Phys.\ Commun.\  {\bf 124}, 76 (2000);
  %%CITATION = HEP-PH/9812320;%%
  %``The Masses of the neutral CP - even Higgs bosons in the MSSM: Accurate analysis at the two loop level,''
  Eur.\ Phys.\ J.\ C {\bf 9}, 343 (1999).
  %%CITATION = HEP-PH/9812472;%%



%\cite{Chattopadhyay:2014gfa}
\bibitem{mssm-color-breaking}
  U.~Chattopadhyay and A.~Dey,
  %``Exploring MSSM for Charge and Color Breaking and Other Constraints in the Context of Higgs@125 GeV,''
  JHEP {\bf 1411}, 161 (2014).
  %%CITATION = doi:10.1007/JHEP11(2014)161;%%
  %20 citations counted in INSPIRE as of 01 Aug 2016



%\cite{Agashe:2014kda}
\bibitem{pdg}
  K.~A.~Olive {\it et al.} [Particle Data Group Collaboration],
  %``Review of Particle Physics,''
  Chin.\ Phys.\ C {\bf 38}, 090001 (2014).
  %%CITATION = doi:10.1088/1674-1137/38/9/090001;%%
  %4452 citations counted in INSPIRE as of 01 Aug 2016

\bibitem{higgsbounds}
  P.~Bechtle {\it et al.},
  %``HiggsBounds 2.0.0: Confronting Neutral and Charged Higgs Sector Predictions with Exclusion Bounds from LEP and the Tevatron,''
  Comput.\ Phys.\ Commun.\  {\bf 182}, 2605 (2011);
  %%CITATION = ARXIV:1102.1898;%%
  %``HiggsBounds: Confronting Arbitrary Higgs Sectors with Exclusion Bounds from LEP and the Tevatron,''
  Comput.\ Phys.\ Commun.\  {\bf 181}, 138 (2010).
  %%CITATION = ARXIV:0811.4169;%%

%\cite{Calibbi:2014lga}
\bibitem{Calibbi:2014lga}
  L.~Calibbi, J.~M.~Lindert, T.~Ota and Y.~Takanishi,
  %``LHC Tests of Light Neutralino Dark Matter without Light Sfermions,''
  JHEP {\bf 1411}, 106 (2014).
  %%CITATION = doi:10.1007/JHEP11(2014)106;%%
  %19 citations counted in INSPIRE as of 01 Aug 2016

    \bibitem{g2-Binosmuon}
%\cite{Endo:2013lva}
  M.~Endo, K.~Hamaguchi, T.~Kitahara and T.~Yoshinaga,
  %``Probing Bino contribution to muon $g - 2$,''
  JHEP {\bf 1311}, 013 (2013)
  doi:10.1007/JHEP11(2013)013
  [arXiv:1309.3065 [hep-ph]].
  %%CITATION = doi:10.1007/JHEP11(2013)013;%%
  %38 citations counted in INSPIRE as of 14 Feb 2017

\bibitem{nath}
R. Arnowitt and P. Nath, Phys. Rev. D 46, 3981 (1992).

%\cite{ArkaniHamed:2006mb}
\bibitem{ArkaniHamed:2006mb}
  N.~Arkani-Hamed, A.~Delgado and G.~F.~Giudice,
  %``The Well-tempered neutralino,''
  Nucl.\ Phys.\ B {\bf 741}, 108 (2006)
  doi:10.1016/j.nuclphysb.2006.02.010
  [hep-ph/0601041].
  %%CITATION = doi:10.1016/j.nuclphysb.2006.02.010;%%
  %288 citations counted in INSPIRE as of 01 Aug 2016

\bibitem{micromega}
  G.~Belanger {\it et al.},
  %``Indirect search for dark matter with micrOMEGAs2.4,''
  Comput.\ Phys.\ Commun.\  {\bf 182}, 842 (2011).
  %%CITATION = ARXIV:1004.1092;%%


\bibitem{axion}
H. Baer, A. Lessa, S. Rajagopalan and W. Sreethawong,
JCAP {\bf 1106} (2011) 031.


%\cite{Ade:2015xua}
\bibitem{planck}
  P.~A.~R.~Ade {\it et al.} [Planck Collaboration],
  %``Planck 2015 results. XIII. Cosmological parameters,''
  arXiv:1502.01589 [astro-ph.CO].
  %%CITATION = ARXIV:1502.01589;%%
  %1979 citations counted in INSPIRE as of 01 Aug 2016


%\cite{Akerib:2013tjd}
\bibitem{Akerib:2013tjd}
  D.~S.~Akerib {\it et al.} [LUX Collaboration],
  %``First results from the LUX dark matter experiment at the Sanford Underground Research Facility,''
  Phys.\ Rev.\ Lett.\  {\bf 112} (2014) 091303.
  %%CITATION = doi:10.1103/PhysRevLett.112.091303;%%
  %1324 citations counted in INSPIRE as of 01 Aug 2016

\bibitem{pandaX}
%\cite{Tan:2016zwf}
  A.~Tan {\it et al.} [PandaX-II Collaboration],
  %``Dark Matter Results from First 98.7 Days of Data from the PandaX-II Experiment,''
  Phys.\ Rev.\ Lett.\  {\bf 117}, no. 12, 121303 (2016)
  doi:10.1103/PhysRevLett.117.121303
  [arXiv:1607.07400 [hep-ex]].
  %%CITATION = doi:10.1103/PhysRevLett.117.121303;%%
  %68 citations counted in INSPIRE as of 21 Nov 2016

\bibitem{lux}
%\cite{Akerib:2016vxi}
  D.~S.~Akerib {\it et al.},
  %``Results from a search for dark matter in the complete LUX exposure,''
  arXiv:1608.07648 [astro-ph.CO].
  %%CITATION = ARXIV:1608.07648;%%
  %70 citations counted in INSPIRE as of 21 Nov 2016

%\cite{Cheung:2012qy}
\bibitem{Cheung:2012qy}
  C.~Cheung, L.~J.~Hall, D.~Pinner and J.~T.~Ruderman,
  %``Prospects and Blind Spots for Neutralino Dark Matter,''
  JHEP {\bf 1305}, 100 (2013)
  doi:10.1007/JHEP05(2013)100
  [arXiv:1211.4873 [hep-ph]].
  %%CITATION = doi:10.1007/JHEP05(2013)100;%%
  %105 citations counted in INSPIRE as of 14 Feb 2017

%\cite{Han:2016qtc}
\bibitem{Han:2016qtc}
  T.~Han, F.~Kling, S.~Su and Y.~Wu,
  %``Unblind the Dark Matter Blind Spots,''
  %Submitted to: JHEP
  [arXiv:1612.02387 [hep-ph]].
  %%CITATION = ARXIV:1612.02387;%%
  %4 citations counted in INSPIRE as of 14 Feb 2017

\bibitem{xenon1t}
XENON1T Collaboration, E. Aprile, {\it et al.}, arXiv:1206.6288.

%\cite{Porod:2003um}
\bibitem{spheno}
  W.~Porod,
  %``SPheno, a program for calculating supersymmetric spectra, SUSY particle decays and SUSY particle production at e+ e- colliders,''
  Comput.\ Phys.\ Commun.\  {\bf 153}, 275 (2003)
  doi:10.1016/S0010-4655(03)00222-4
  [hep-ph/0301101].
  %%CITATION = doi:10.1016/S0010-4655(03)00222-4;%%
  %654 citations counted in INSPIRE as of 01 Aug 2016


%\cite{Alwall:2014hca}
\bibitem{mad5}
  J.~Alwall {\it et al.},
  %``The automated computation of tree-level and next-to-leading order differential cross sections, and their matching to parton shower simulations,''
  JHEP {\bf 1407}, 079 (2014).
  %%CITATION = doi:10.1007/JHEP07(2014)079;%%
  %1099 citations counted in INSPIRE as of 21 Jul 2016

\bibitem{pythia}
  T.~Sjostrand, S.~Mrenna and P.~Z.~Skands,
  %``PYTHIA 6.4 Physics and Manual,''
  JHEP {\bf 0605}, 026 (2006).
  %%CITATION = HEP-PH/0603175;%%

\bibitem{delphes}
  J.~de Favereau,  {\it et al.},
  %``DELPHES 3, A modular framework for fast simulation of a generic collider experiment,''
arXiv:1307.6346 [hep-ex].


\bibitem{fastjet}
  M.~Cacciari, G.~P.~Salam and G.~Soyez,
  %``FastJet User Manual,''
  Eur.\ Phys.\ J.\ C {\bf 72}, 1896 (2012)
  [arXiv:1111.6097 [hep-ph]].
  %%CITATION = ARXIV:1111.6097;%%
  %922 citations counted in INSPIRE as of 24 Jul 2015

\bibitem{anti-kt}
  M.~Cacciari, G.~P.~Salam and G.~Soyez,
  %``The Anti-k(t) jet clustering algorithm,''
  JHEP {\bf 0804}, 063 (2008).
  %%CITATION = ARXIV:0802.1189;%%


\bibitem{atlas-13tev-ewkino-1}
ATLAS-CONF-2016-075

\bibitem{cms-13tev-ewkino-1}
CMS-PAS-SUS-16-024.


\bibitem{cms-13tev-ewkino-2}
CMS-PAS-SUS-16-022

%\cite{Aad:2016tuk}
\bibitem{atlas-13tev-ewkino-2}
  G.~Aad {\it et al.} [ATLAS Collaboration],
  %``Search for supersymmetry at $\sqrt{s}=13$  TeV in final states with jets and two same-sign leptons or three leptons with the ATLAS detector,''
  Eur.\ Phys.\ J.\ C {\bf 76}, no. 5, 259 (2016)
  doi:10.1140/epjc/s10052-016-4095-8
  [arXiv:1602.09058 [hep-ex]].
  %%CITATION = doi:10.1140/epjc/s10052-016-4095-8;%%
  %25 citations counted in INSPIRE as of 02 Nov 2016

\bibitem{checkmate}
  M.~Drees {\it et al.},
  %``CheckMATE: Confronting your Favourite New Physics Model with LHC Data,''
  Comput.\ Phys.\ Commun.\  {\bf 187}, 227 (2014).
  %%CITATION = ARXIV:1312.2591;%%
  %52 citations counted in INSPIRE as of 24 Jul 2015

%\cite{Beenakker:1996ed}
\bibitem{prospino}
  W.~Beenakker, R.~Hopker and M.~Spira,
  %``PROSPINO: A Program for the production of supersymmetric particles in next-to-leading order QCD,''
  hep-ph/9611232.
  %%CITATION = HEP-PH/9611232;%%
  %485 citations counted in INSPIRE as of 01 Aug 2016


%\cite{Barr:2015eva}
\bibitem{lhc-compressed-1}
  A.~Barr and J.~Scoville,
  %``A boost for the EW SUSY hunt: monojet-like search for compressed sleptons at LHC14 with 100 fb$^{−1}$,''
  JHEP {\bf 1504}, 147 (2015).
  %%CITATION = doi:10.1007/JHEP04(2015)147;%%
  %6 citations counted in INSPIRE as of 01 Aug 2016

%\cite{Schwaller:2013baa}
\bibitem{lhc-compressed-2}
  P.~Schwaller and J.~Zurita,
  %``Compressed electroweakino spectra at the LHC,''
  JHEP {\bf 1403}, 060 (2014).
  %%CITATION = doi:10.1007/JHEP03(2014)060;%%
  %64 citations counted in INSPIRE as of 01 Aug 2016

%\cite{Han:2013usa}
\bibitem{lhc-compressed-3}
  C.~Han, A.~Kobakhidze, N.~Liu, A.~Saavedra, L.~Wu and J.~M.~Yang,
  %``Probing Light Higgsinos in Natural SUSY from Monojet Signals at the LHC,''
  JHEP {\bf 1402}, 049 (2014).
  %%CITATION = doi:10.1007/JHEP02(2014)049;%%
  %73 citations counted in INSPIRE as of 01 Aug 2016

%\cite{Baer:2014cua}
\bibitem{lhc-compressed-4}
  H.~Baer, A.~Mustafayev and X.~Tata,
  %``Monojets and mono-photons from light higgsino pair production at LHC14,''
  Phys.\ Rev.\ D {\bf 89}, no. 5, 055007 (2014).
  %%CITATION = doi:10.1103/PhysRevD.89.055007;%%
  %47 citations counted in INSPIRE as of 01 Aug 2016

%\cite{Han:2014xoa}
\bibitem{lhc-compressed-4.5}
  C.~Han, L.~Wu, J.~M.~Yang, M.~Zhang and Y.~Zhang,
  %``New approach for detecting a compressed bino/wino at the LHC,''
  Phys.\ Rev.\ D {\bf 91}, 055030 (2015).
  %%CITATION = doi:10.1103/PhysRevD.91.055030;%%
  %26 citations counted in INSPIRE as of 04 Aug 2016

%\cite{Dutta:2014jda}
\bibitem{lhc-compressed-5}
  B.~Dutta {\it et al.},
  %``Probing Compressed Sleptons at the LHC using Vector Boson Fusion Processes,''
  Phys.\ Rev.\ D {\bf 91}, no. 5, 055025 (2015).
  %%CITATION = doi:10.1103/PhysRevD.91.055025;%%
  %10 citations counted in INSPIRE as of 01 Aug 2016


%\cite{Dutta:2013gga}
\bibitem{lhc-compressed-6}
  B.~Dutta {\it et al.},
  %``Probing compressed top squark scenarios at the LHC at 14 TeV,''
  Phys.\ Rev.\ D {\bf 90}, no. 9, 095022 (2014).
  %%CITATION = doi:10.1103/PhysRevD.90.095022;%%
  %25 citations counted in INSPIRE as of 01 Aug 2016

%\cite{Arkani-Hamed:2015vfh}
\bibitem{Arkani-Hamed:2015vfh}
  N.~Arkani-Hamed, T.~Han, M.~Mangano and L.~T.~Wang,
  %``Physics Opportunities of a 100 TeV Proton-Proton Collider,''
  arXiv:1511.06495 [hep-ph].
  %%CITATION = ARXIV:1511.06495;%%
  %41 citations counted in INSPIRE as of 01 Aug 2016


%\cite{Low:2014cba}
\bibitem{100tev-ewkino-4}
  M.~Low and L.~T.~Wang,
  %``Neutralino dark matter at 14 TeV and 100 TeV,''
  JHEP {\bf 1408}, 161 (2014).
  %%CITATION = doi:10.1007/JHEP08(2014)161;%%
  %91 citations counted in INSPIRE as of 01 Aug 2016

\end{thebibliography}
\end{document}